\documentclass[acmsmall,screen]{acmart}
\usepackage{enumitem}
\usepackage{graphicx}
\usepackage{adjustbox}
\usepackage{multirow}
\usepackage{subcaption}
\usepackage[most]{tcolorbox}
\usepackage{xcolor}        
\usepackage{colortbl}
\usepackage{amsmath}    
\usepackage{pgfplots}
\pgfplotsset{compat=1.18}
\usepackage[]{collab}
\usepackage[utf8]{inputenc}
\usepackage{algorithm}
\usepackage{algorithmic}

\usepackage{amssymb}
\usepackage{pgfplots}

\AtBeginDocument{%
  }

\collabAuthor{yt}{red}{Yintong Huo}

\begin{document}

\title{Small Is Beautiful: A Practical and Efficient Log Parsing Framework}

\author{Minxing Wang}
\orcid{0009-0006-5741-9920}
\affiliation{%
  \institution{Singapore Management University}
  \city{Singapore}
  \country{Singapore}
}
\email{mxwang@smu.edu.sg}

\author{Yintong Huo}
\orcid{0009-0006-8798-5667}
\affiliation{%
  \institution{Singapore Management University}
  \city{Singapore}
  \country{Singapore}
}
\email{ythuo@smu.edu.sg}

\begin{abstract}
  Log parsing serves as the fundamental step in log analysis, splitting logs into constant templates and dynamic variables. While recent semantic-based parsers leveraging LLM have shown superior generalizability over prior syntax-based methods, their effectiveness is critically dependent on the scale of the underlying model. This dependency results in a significant performance collapse when using smaller, more practical LLMs, thereby creating a major barrier to real-world adoption where data privacy and computational constraints necessitate the use of succinct and resource-efficient models.

In a typical semantic parsing pipeline, the parsing cache is a critical component that stores the set of observed templates to quickly route incoming logs. The design of this cache is therefore paramount to the parser's overall effectiveness. Motivated by such, we improve parsing accuracy from two insights: 1) designing a more flexible cache updating strategy that can rectify prior errors, and 2) including an explicit validation process to proofread templates before they are added to the cache, preventing error injection. In particular, we propose EFParser, an unsupervised LLM-based log parser, including template extraction, template correction, and validated templates. To mitigate the impact of degraded capabilities in smaller LLMs, we designed a dual cache with an adaptive updating mechanism. When the LLM generates a new template, this module determines if it is a novel pattern or a variation of an existing one. If it's a variation, it merges the templates, thereby maintaining consistency and correcting the cache. Furthermore, we integrate a correction module that acts as a gatekeeper, validating and refining every LLM-generated template to ensure only high-quality, accurate patterns are cached. Evaluation on public large-scale datasets demonstrates that EFParser outperforms all baseline methods by an average of 12.5\% across all evaluation metrics when running on smaller LLMs, with performance that even exceeds some baseline methods using large-scale LLMs, highlighting the advantages of systematic architectural design. Moreover, despite the additional processing procedures, the average processing time remains shorter than most semantic-based baselines. The superior performance on smaller LLMs combined with computational efficiency demonstrates that EFParser has significant potential for real-world deployment.
\end{abstract}



\maketitle

\section{Introduction}
 Log messages, generated by various components such as the operating system, software applications, and hardware devices, record system action trails and historical runtime states, providing valuable data for enhancing system reliability ~\cite{he2022empirical, li2024go, li2023exploring}. Currently, numerous works are dedicated to log analysis, including anomaly detection ~\cite{guo2021logbert, he2016experience, zhang2022deeptralog, zhang2019robust}, root cause analysis ~\cite{lu2017log, zhang2019robust, wang2020root}, and failure troubleshooting ~\cite{he2018identifying, xu2009detecting, chen2021pathidea}. Among all, log parsing serves as the first and fundamental step in these log analysis works, which splits a log message into two parts: 1) \textit{log templates (i.e., constants):} static parts that provide a high-level overview of the event; 2) \textit{log parameters (i.e., variables):} dynamic parts that can only be determined at runtime and provide details of the event. For example, a log message like ``\texttt{Scheduled snapshot period at 10 seconds.}'' can be parsed into the template ``\texttt{Scheduled snapshot period at <$*$> seconds.}" and the parameter \texttt{10}.

Existing log parsing techniques can be broadly categorized into syntax-based and semantic-based approaches. Syntax-based log parsers utilize heuristics or statistical features (e.g., token count, frequency, and position) to identify the constant templates ~\cite{he2017drain, vaarandi2015logcluster, xu2023hue}. While effective in simple data, they often struggle with the increased diversity and complexity in contemporary logs. More recently, semantic-based log parsers leveraging Large Language Models (LLMs) have emerged, demonstrating superior performance by understanding log semantics beyond mere syntactic structure ~\cite{xu2024divlog, jiang2024lilac, xiao2024free, huang2025no}. By exploiting natural language understanding and generation capabilities, LLM-based parsers achieve greater generalizability and robustness across a wide range of log data. 


However, the practical deployment of these advanced parsers is hindered by a critical challenge: their performance is highly contingent on large-scale models (e.g., GPT-3.5), which are often infeasible for industrial local deployment due to data privacy and operational costs ~\cite{das2025security, pan2020privacy}. While these parsers can be adapted to smaller, locally-hosted models, this transition results in a significant degradation of accuracy.  We analyze the reasons as follows.

First, the performance degradation originates in the parser’s ``control center'': the parsing cache. This component is responsible for matching incoming logs to existing templates or routing to generate new ones.
However, current caches are designed with a prefix tree-based structure, which restricts template retrieval using single-direction traversal.
Moreover, their rule-based updating strategies often fail to merge semantically similar logs. 
Consequently, the cache incorrectly generates redundant templates instead of unifying related ones. Without an effective strategy to rectify these errors, template mistakes accumulate throughout the parsing process. This creates a cascading failure effect, where erroneous templates propagate and affect subsequent logs, severely compromising the parsing performance.

Second, current parsers overlook the critical need for template validation before pushing it into cache. While large models generate reliable outputs, smaller LLMs tend to trigger more hallucinations. Specifically, they have a significantly higher propensity to misclassify constants and variables within a log template. By accepting these problematic templates without a validation step, parsers allow errors to compound. This introduces another layer of cascading failures, further degrading parsing accuracy over time.

To address the above shortcomings of existing unsupervised log parsers and adapt them for smaller LLMs, we propose EFParser, an unsupervised log parsing framework leveraging small LLMs. EFParser consists of three main components: the reference log exemplar selection module, the dual parsing cache, and the correction module. The exemplar selection module enables accurate unsupervised log parsing by selecting logs that balance similarity and diversity with the target log entry, allowing LLMs to distinguish constants from variables through positional token analysis and semantic understanding. The dual parsing cache design incorporates a tree-and-bucket-based cache architecture with an adaptive updating mechanism. Upon receiving newly generated templates, the system identifies cached templates that exhibit structural and semantic similarities to the incoming templates. For such related templates, the parsing cache employs either rule-based heuristics or LLM-assisted methods to perform selective or comprehensive template merging, thereby enhancing template quality and consolidating semantically equivalent patterns. The correction module is specifically designed to address the inherent limitations of smaller LLMs in hallucination and distinguishing between constants and variables within complex log structures. Through targeted correction strategies, this module systematically rectifies faulty structures and misclassifications, improving overall parsing accuracy.

We have conducted a comprehensive evaluation on the 14 large-scale datasets in Loghub-2.0 from LogPAI. Our experimental results demonstrate that EFParser outperforms all baseline unsupervised parsers when run on an 8B model. Specifically, EFParser surpasses LILAC w/o ICL by a significant margin, achieving an F1 score that is 23.8\% higher for grouping accuracy and 50.6\% higher for template accuracy. Compared to LUNAR, EFParser also shows superior performance, with an F1 score that is 2.9\% higher for grouping accuracy and 7.7\% higher for template accuracy, respectively. Furthermore, EFParser demonstrates strong robustness across all datasets and maintains high performance when transferred to other small models. EFParser's superior performance and generalizability prove its potential for large-scale adoption and practical use within industry.

To sum up, the main contributions of this paper are threefold:
\begin{itemize}
    \item To our best knowledge, we propose EFParser, the first unsupervised practical log parsing framework for small language models (<10B) that achieves performance competitive with that of LLMs. 
    \item We introduce two novel modules to address small language models' limitations in complex log parsing scenarios: 1) dual parsing cache with an adaptive updating mechanism triggered by newly incoming logs, and 2) a correction module for log templates generated by LLM. 
    \item We conduct a comprehensive evaluation on the public large-scale Loghub-2.0 dataset. Results demonstrate that EFParser outperforms all unsupervised state-of-the-art methods in accuracy while maintaining strong robustness.
\end{itemize}

\section{Background and Motivation}
In this section, we describe the typical workflow of semantic parsers as background, and introduce three limitations within them.
\subsection{Typical Workflow of Semantic Parsers}

Typical LLM-based semantic-based log parsers consist of three core components: \textit{a parsing cache}, \textit{a demonstration selector}, and \textit{an LLM-powered template extraction module}. Through the integration of these three modules, semantic-based log parsers achieve high template extraction accuracy while maintaining processing efficiency. These parsers like LILAC ~\cite{jiang2024lilac}, LogBatcher ~\cite{xiao2024free}, and LUNAR ~\cite{huang2025no} have exhibited exceptional performance on the comprehensive Loghub-2.0 benchmark, consistently surpassing syntactic-based baseline methods ~\cite{jiang2024large}.

\begin{figure}
    \vspace{2mm}
    \centering
    \includegraphics[width=\linewidth]{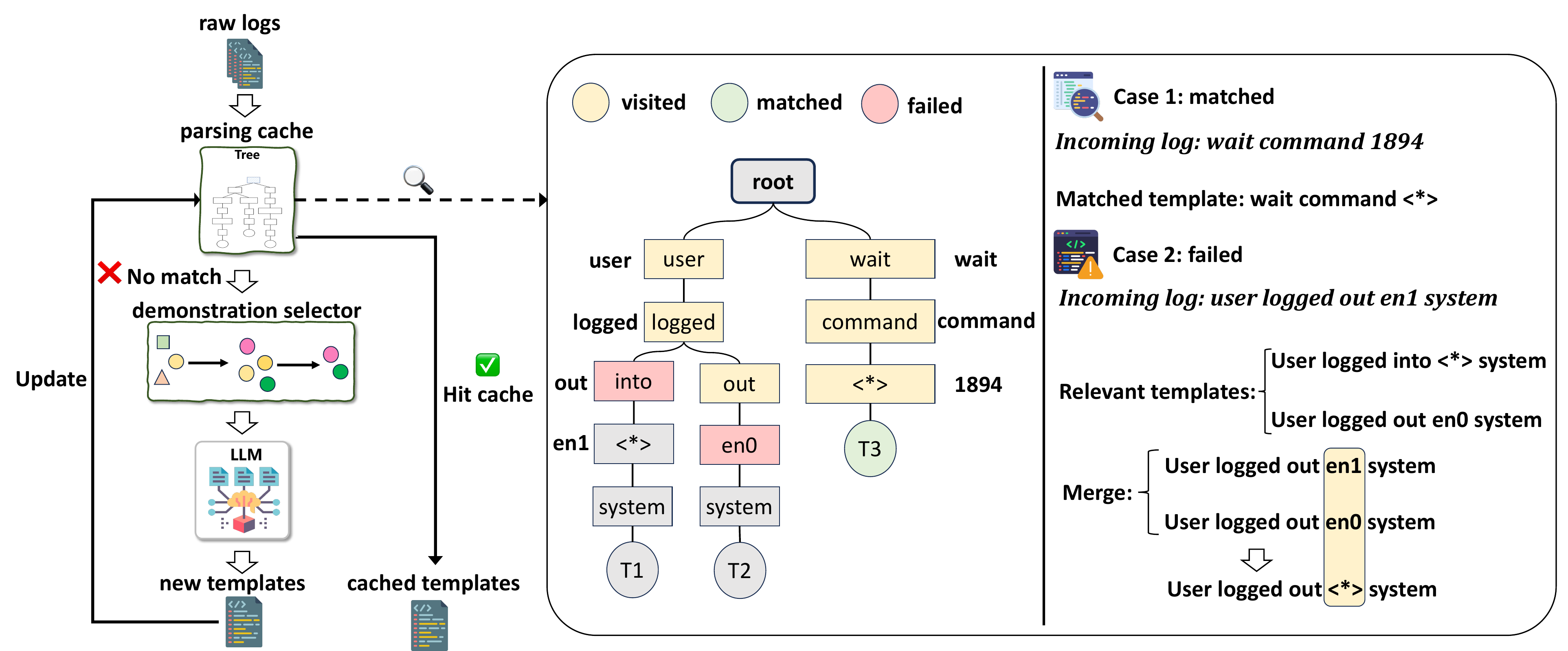}
    \caption{General workflow of semantic-based log parsing.}
    \label{fig:semantic-based_parser_workflow}
\end{figure}

The workflow of LLM-based parsers is shown in Fig.~\ref{fig:semantic-based_parser_workflow}. When a new log entry arrives, it first passes through the parsing cache, where the parser determines whether this log exhibits a pattern that matches existing templates. The parsing cache serves for storing templates generated by LLMs, and these cached templates are commonly structured in a \textit{tree} structure. If a match is found (Case 1 in the figure), the corresponding template is returned directly without requiring further queries to the LLM, thereby reducing the number of LLM token consumption and ensuring parsing consistency ~\cite{jiang2024lilac}. 


Case 2 in Fig. \ref{fig:semantic-based_parser_workflow} illustrates the template update process. Consider an incoming log entry "user logged out $en1$ system" for which no matching cached template exists. In such cases, the system retrieves all cached templates containing the failed parsing nodes (specifically "into" and "$en0$" in this example) and identifies the most similar template among them. This candidate template is then compared with the new template generated by the LLM. If they are determined to represent the same underlying template structure, the system merges the differing tokens at their corresponding positions to create a unified template. In this example, the original cached template "user logged out $en0$ system" misclassified "$en0$" as a constant; this error is corrected following the template update process.

If the parsing cache contains no matching pattern for the incoming log, the parser then directs to an LLM to generate a template. Contemporary semantic parsers differ in how they select demonstrations: Supervised methods ~\cite{jiang2024lilac} retrieve similar, labeled templates from the dataset to guide the LLM, while unsupervised ones ~\cite{huang2025no} sample representative logs that balance commonality and variability with the target log to provide context. After selecting the demonstrations, the query module prompts LLM to generate a template, which is subsequently updated into the parsing cache for reuse.


\subsection{Limitations}
While existing semantic parsers focus on curating LLM prompts with selected demonstrations, they overlook issues raised by their underlying architectural design. 
We argue that semantic parsers' effectiveness strongly relies on the quality of the internal caching strategy and LLM-powered template generation. In this section, we identify three limitations that might hinder their practicality in real-world applications.



\subsubsection{Inflexible Tree-Based Cache Updating Strategy}
From the previous section, we recognize that the parsing cache is central to the parser: it routes incoming logs and acts as the only reference for direct parsing (hitting the cache). Consequently, 
the cache’s self-correcting ability would directly determine overall parsing performance, which relies on an effective \textit{cache updating} strategy.
When a cache miss occurs, updating proceeds in two phases: 1) retrieving relevant templates from cache, and 2) merging/inserting the current templates. 
Prior work uses \textit{fixed rule-based} updates that compare newly generated LLM templates against cached ones to merge and correct prior errors. While handcrafted rules are effective for simple datasets, they require strict conditions and degrade in performance for complex scenarios.


The \textbf{first limitation} is the \textit{restricted template retrieval} in a tree-based cache. These log parsers use a prefix tree-based cache to find matching templates for incoming logs.
To search for relevant templates for the target log, existing parsers rely on a prefix tree structure that begins matching tokens from the start of a log entry, and retrieve all associated paths (T1 and T2 in Fig.~\ref{fig:semantic-based_parser_workflow}). This procedure inherently assumes the first token is always a constant rather than a variable.
In practice, however, this assumption is often incorrect. 
If a cached template mistakenly marks its first token as a constant, any incoming log will fail to match at the root of the tree. Consequently, these cached template mistakes are immediately excluded from actual relevant templates and cannot be corrected after, thereby limiting the effectiveness of the updating mechanism.

This problem frequently appears in datasets where many templates start with "<$*$>". For example, in the Loghub-2.0 Linux dataset, about 10\% of templates begin this way and are affected by this limitation. Given the central role of the parsing cache, this issue could substantially degrade parsing performance, making this a critical issue for mitigation.

The \textbf{second limitation} is the \textit{rigidity of rule-based template updating}.
These strategies can only merge two templates if they have the exact same token length, proceeding with a strict, position-by-position comparison that replaces any mismatch with a placeholder ``<$*$>''. This will cause two serious flaws. First, it fails to merge templates with different token counts, even when they represent the same underlying log pattern. Taking Figure \ref{fig:merge_a} as an example, a template that incorrectly splits a variable into two separate tokens (e.g., ``\texttt{Mon}'' and ``<$*$>'') cannot be corrected with a more accurate template to treat them as a single variable (``\texttt{<$*$>}'') later.
Second, it cannot perform partial or semantic updating. Often, two templates are highly similar but semantically different. An effective strategy would adaptively merge the common variable parts (like a hostname) while preserving the distinct, constant parts (like specific error messages) that contain critical anomaly information.



\begin{tcolorbox}[boxsep=1pt,left=2pt,right=2pt,top=3pt,bottom=2pt,width=\linewidth,colback=white!90!black,boxrule=0pt, colbacktitle=white!,toptitle=2pt,bottomtitle=1pt,opacitybacktitle=0]
\textbf{Observation 1.} Current methods fail to rectify cached template errors and would cascadingly degrade the effectiveness of parsing, highlighting the need for a more flexible and adaptive cache updating strategy.
\end{tcolorbox}



\begin{figure}[h!]
    \centering

    \begin{subfigure}[t]{0.45\textwidth}
        \footnotesize
        \begin{tcolorbox}[
            colback=white,   
            colframe=black, 
            rounded corners,        
            boxrule=1pt,
            height=11em, boxsep=1pt,left=4pt,right=4pt,top=5pt,bottom=5pt
        ]
            \texttt{$L_1$: connection from 210.202.115.220 at \textcolor{red}{Mon Aug 9 09:12:50 2005}} \\[1.5ex]
            \texttt{$L_2$: connection from 261.218.671.604 at \textcolor{red}{Tue Oct 13 02:01:27 2005}} \\[1.5ex]
            \texttt{$T_1$: connection from <$*$> at \textcolor{red}{Mon <$*$>}} \\[1.5ex]
            \texttt{$T_2$: connection from <$*$> at \textcolor{red}{<$*$>}}
        \end{tcolorbox}
        \caption{Templates with Different Lengths Requiring Merge}
        \label{fig:merge_a}
    \end{subfigure}
    \hfill
    \begin{subfigure}[t]{0.5\textwidth}
    \footnotesize
        \begin{tcolorbox}[
            colback=white,
            colframe=black,
            rounded corners,
            boxrule=1pt,
            height=11em, boxsep=1pt,left=4pt,right=4pt,top=5pt,bottom=5pt
        ]
            \vspace{1ex}
            \texttt{$T_1$: ERROR: Database connection \textcolor{orange}{failed} - Host: \textcolor{red}{db-primary}} \\[0ex]
            \texttt{$T_2$: ERROR: Database connection \textcolor{orange}{refused} - Host: \textcolor{red}{db-cache}}\\[0ex]
            \texttt{(Ground-truth) $T_1'$: ERROR: Database connection \textcolor{orange}{failed} - Host: \textcolor{red}{<*>}}\\[0ex]
            \texttt{(Ground-truth) $T_2'$: ERROR: Database connection \textcolor{orange}{refused} - Host: \textcolor{red}{<*>}}
        \end{tcolorbox}
        \caption{Templates Requiring Partial Merge}
        \label{fig:merge_b}
    \end{subfigure}
    \caption{Examples of Rule-based Merging Failures}
\end{figure}

\subsubsection{Performance Dependence on LLM Backbones.}
The \textbf{third limitation} is the over-reliance on LLM scales. While LLMs provide a strong backbone for semantic understanding, their practical deployment is constrained by several factors~\cite{hou2025unveiling}. Privacy and security protocols often prohibit using third-party commercial LLMs ~\cite{carlini2021extracting}. Furthermore, the substantial computational and energy demands of LLMs are major barriers to widespread adoption, raising ``Green AI'' concerns ~\cite{li2024sprout, shi2025efficient}. 
To this end, there is some ongoing work to downscale LLMs to preserve performance while largely reducing resource demands ~\cite{goel2025position, belcak2025small}.
Therefore, we argue that attention should shift some attention toward optimizing semantic parsers for more resource‑efficient, smaller LLMs.

\begin{figure}
    \vspace{2mm}
    \centering
    \includegraphics[width=0.85\linewidth]{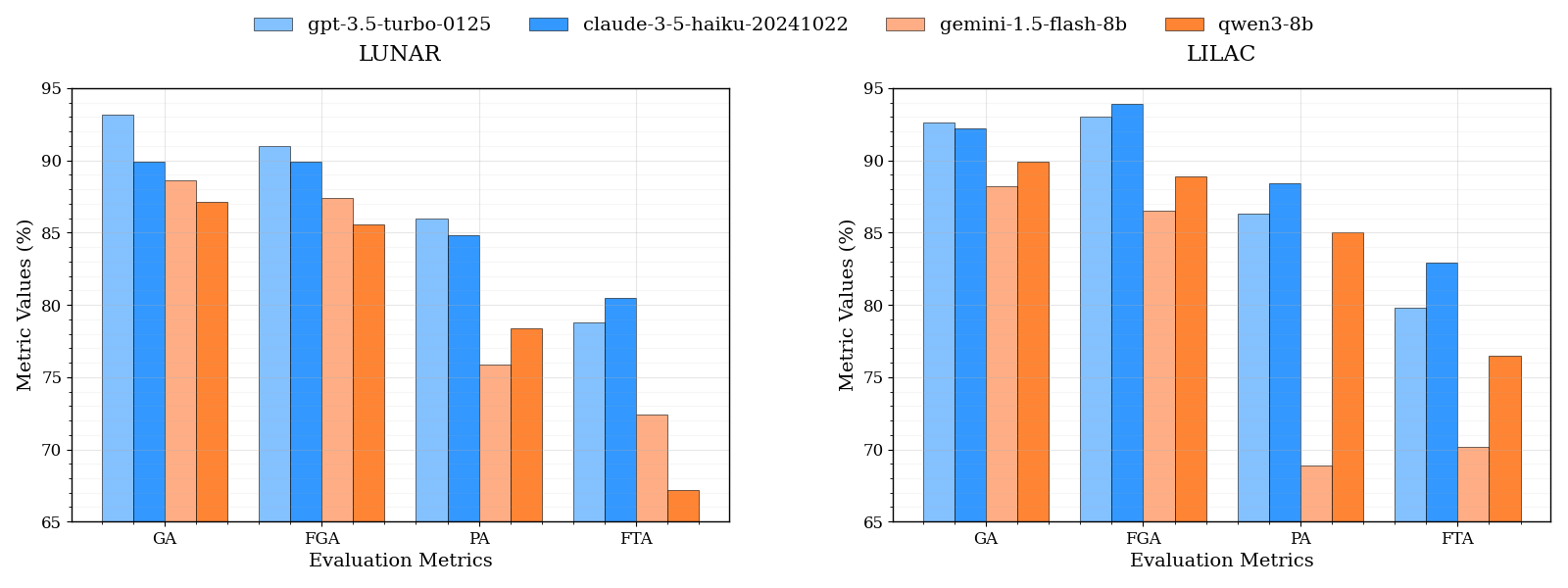}
    \caption{Performance of Semantic Log Parsers Across Different LLM Scales}
    \label{fig:performance_comparison_large_small}
\end{figure}

Our preliminary analysis reveals that parser performance is highly contingent on model size and degrades notably when using smaller models. This is because LLM, as the template extraction module, encounters more hallucinations with its smaller size: a hallucinated template can produce cascading failures for subsequent logs.
Figure \ref{fig:performance_comparison_large_small} compares performance for LILAC ~\cite{jiang2024lilac} and LUNAR ~\cite{huang2025no} across different LLM scales. When these methods are run with 8b-size LLMs, F1 scores for grouping accuracy and template accuracy decrease by 5.3\% and 11.1\%, respectively, underscoring the need to design parsers that are robust on smaller backbones.
The further analysis identifies that the performance discrepancy stems from their limited capacity to attend to all variables in lengthy and structurally complex log entries. For instance, when LUNAR and LILAC w/o ICL are transferred from GPT-3.5 to Gemini-1.5-flash-8b, their performance on the complex OpenStack dataset decreases by 30.2\% and 10.1\% in terms of GA and PA metrics, respectively. In contrast, on simple datasets such as HDFS, performance remains virtually unchanged when the underlying model is switched.

\begin{tcolorbox}[boxsep=1pt,left=2pt,right=2pt,top=3pt,bottom=2pt,width=\linewidth,colback=white!90!black,boxrule=0pt, colbacktitle=white!,toptitle=2pt,bottomtitle=1pt,opacitybacktitle=0]
\textbf{Observation 2.} Existing semantic parsers degrade substantially on smaller models due to their inability to handle complex logs, necessitating a more scalable approach to manage the hallucinated templates.
\end{tcolorbox}

\section{Methodology}
\subsection{Overview}
\begin{figure}[t]
  \vspace{2mm}
  \centering
  \includegraphics[width=1.0\textwidth]{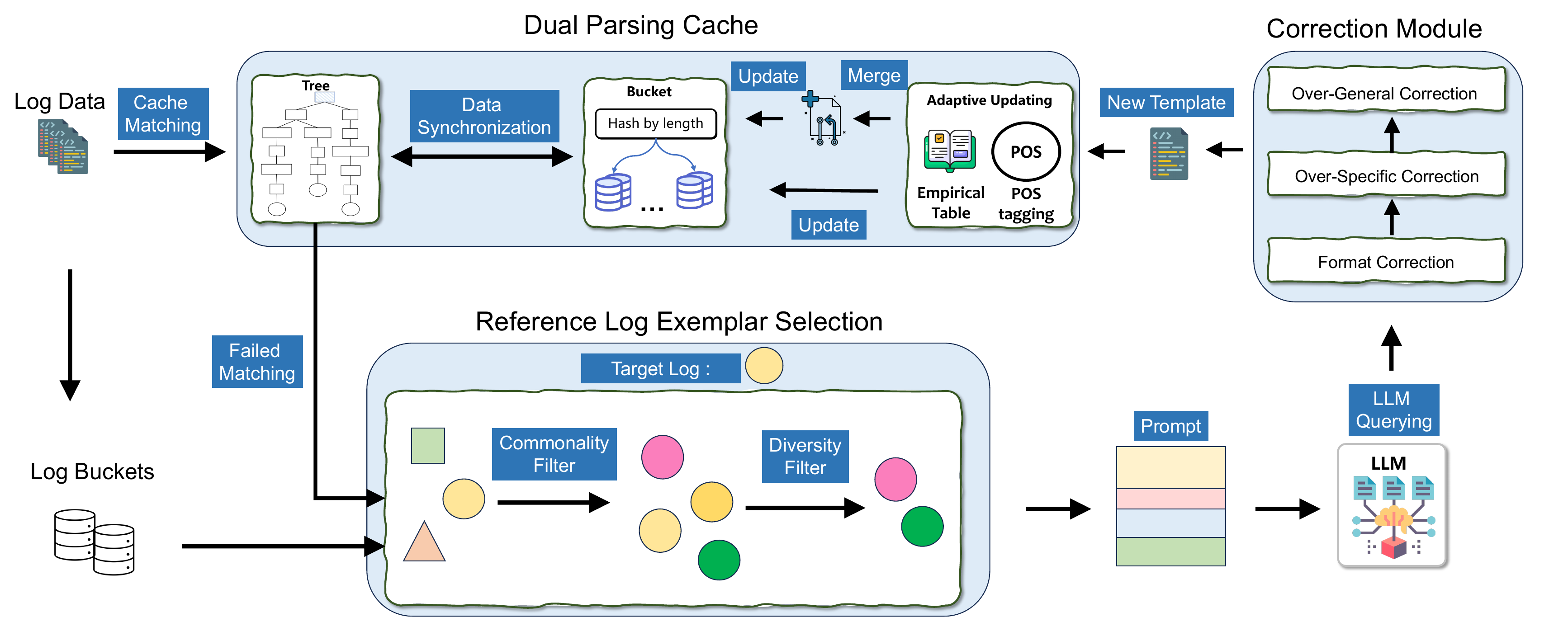}
  \caption{The overall workflow of EFParser}
  \label{fig:workflow_EFParser}
\end{figure}
EFParser comprises three core components: a dual tree-bucket parsing cache (dual cache), a correction module, and a reference log exemplar selector (exemplar selector). To address the limitations of restricted template retrieval, we combine tree and bucket over a dual cache to share templates, enabling a global search for relevant templates while maintaining high retrieval efficiency. 
Additionally, we employ an adaptive template merging approach to update the dual cache. 
To reduce degradation on the smaller models, the correction module targets common hallucination patterns and refines LLM-generated templates.
Finally, the exemplar selection module identifies logs that balance commonalities and differences with the target log, providing demonstrations for template extraction. These three components work collaboratively to deliver effectiveness, even when deployed on smaller LLMs.

As illustrated in Figure \ref{fig:workflow_EFParser}, when a new log arrives, EFParser first queries the tree cache for existing matches. If a match is found, it returns the cached template and ends parsing. 
Otherwise, the exemplar selection module identifies reference logs and prompts the LLM to generate a template for the target log.
The generated template is refined by the correction module to handle potential hallucinations. 
Then, the bucket cache performs a global match of the generated template against existing ones and decides whether an update is needed.
If updating is required, EFParser employs an adaptive method to merge templates based on their syntax information and semantic information. The dual cache is then updated accordingly. 
If no merge is required, it inserts the new template into both caches.
The following sections provide a detailed description of each component of EFParser.

\subsection{Dual Parsing Cache}
\subsubsection{Dual Cache Initialization}

Our parser's central data structure is a dual cache designed to serve two distinct but complementary purposes: rapid template retrieval for known patterns and efficient global search for relevant templates during cache updates (for failed matches). To achieve this, we maintain the same cache content using two synchronized data structures: a prefix tree for fast lookups and a bucket-based index for comprehensive global searches.

The tree-based cache follows the design in prior semantic-based parsers. Its sole function is to provide low-latency, prefix-based matching for incoming logs against existing templates. This structure is good at handling the case where a log matches a known pattern, but it returns an insufficient set of similar templates upon a failed match.

To overcome the search limitations inherent in tree-based structures, we introduce a novel bucket-based cache. This second structure indexes the same set of templates but organizes them differently: it groups templates into buckets based on their token count. 
Under this design, we could use the token length as a key to query relevant templates when a new template is generated (e.g., by the LLM). This method enables global search while constraining the search to a small and relevant subset of the entire template databases.


During cache updates, both data structures are synchronized simultaneously. While maintaining dual data structures introduces a memory overhead, our analysis of large-scale, real-world datasets shows this cost is negligible: even datasets containing tens of millions of logs rarely exceed 5,000 unique templates. 
This modest overhead is a worthwhile trade-off for the gains in search comprehensiveness and accuracy, which directly improve the quality and consistency of the template cache over time. 

\subsubsection{Adaptive Cache Updates}
To maintain an up-to-date cache, our parser employs an adaptive template-updating mechanism.
Given an unseen template, this step determines how to update the existing cache by two steps: (1) retrieving the most relevant template, and (2) deciding whether to merge it with the current cached template or add it as a new, distinct entry.
Unlike prior approaches that depend on a fixed set of heuristics, our method leverages a combination of token-level, syntactic, and semantic similarities to update the cache.

The first retrieval process begins by identifying the most suitable candidate for a potential merge. Specifically, given an unseen template $T_{target}$ with length $l_{target}$, the first step is to select the most relevant template based on its token-level Levenshtein distance, where $ED$ represents the edit distance.
To avoid extensive comparison against all cached templates, we use the bucket cache to perform an efficient, length-based pre-selection. Particularly, 
we construct a set $R$ to include templates $T_{relevant}$, whose length  $l_{relevant}$ falls within a predefined threshold (as defined in following equation).
If the most relevant template in $R$ meets the similarity threshold, we input the new template $T_{target}$ and the most relevant one into the next step.


$$
\left\{
\begin{aligned}\label{equ:length_range}
&ED_{min} = |l_{target} - l_{relevant}| \\
&similarity = 1 - \frac{ED}{\max(l_{target}, l_{relevant})} \\
&threshold \leq similarity
\end{aligned}
\right.
\quad \Rightarrow \quad
\lceil threshold \times l_{target}\rceil \leq l_{relevant} \leq \lfloor \frac{l_{target}}{threshold}\rfloor
\label{eq: relavant}
$$

\begin{figure}[tbp]
    \vspace{2mm}
    \centering
    \includegraphics[width=1.0\linewidth]{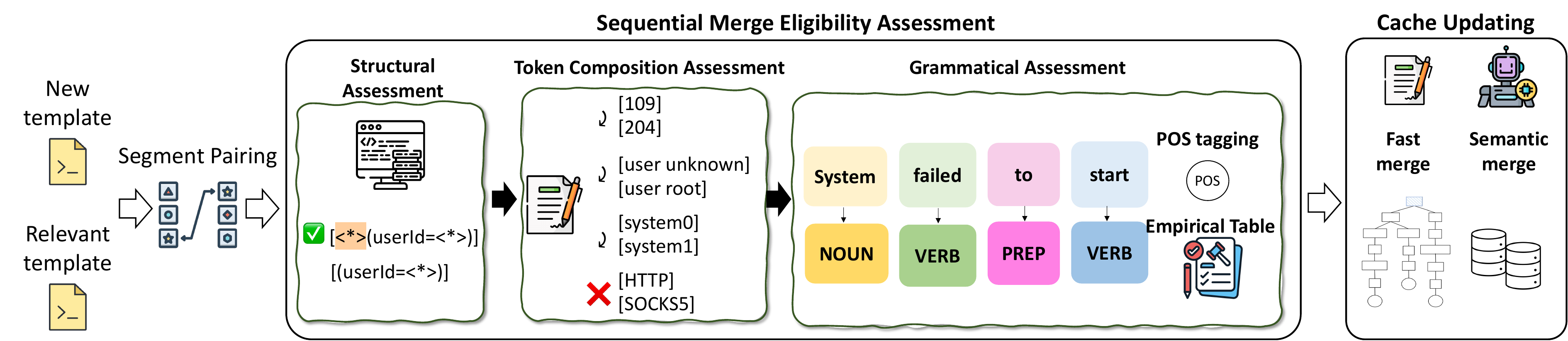}
    \caption{The workflow of cache updates}
    \label{fig:cache_updates}
\end{figure}

In the second step, the two templates will go through a segment pairing to uncover their difference, then apply sequential assessments to decide the update behavior (merge or insert), as shown in Figure \ref{fig:cache_updates}.
To extract the difference between a pair of templates, we apply the Longest Common Subsequence (LCS) algorithm to identify the longest sequence of identical tokens shared by both templates ~\cite{bergroth2000survey, hirschberg1977algorithms}. This common subsequence is then used as a set of delimiters to partition both templates. LCS is known to be resilient to positional shifts, so that these differential parts are not affected by the different lengths of templates.


We then decide if these discrepant segments are able to be merged by a sequential assessment process. The central insight behind our coarse-to-fine process is that not all differences between templates are equally significant. Some are safe to merge (e.g., numeric variables), while others must be preserved to maintain semantic integrity (e.g., verbs). We therefore adopt an adaptive strategy that handles the most common and least risky merges first, progressing to more challenging semantic comparisons only when necessary. Specifically, the sequential checks include three phases, with each acting as a decision gate: 


\begin{enumerate}[leftmargin=14pt, itemsep=0.5em]
    \item Structural assessments. It determines whether two differing segments vary only at  <$*$> positions. If they are identical, the segments are merged immediately without further assessments. This efficiently handles trivial differences, as <$*$> are placeholders for variable content and their positions do not alter the core meaning of the template.
    \item Token composition assessments. If the first check does not result in a merge, this phase ensures that tokens in differing positions share the same data type: alphabetic-only, numeric-only, or alphanumeric. Tokens are considered type-compatible only if they share identical character compositions. Type mismatches will be ruled out, as merging incompatible types (like a timestamp with a user name) would create a template that is too broad to be useful; Otherwise, the process advances to the final rule-based syntactic filtering phase. 
    \item Grammatical assessments. For segments that pass the first two checks, this final phase checks the syntactic roles for each token. It permits merging only if two conditions are met: (a) at least one token is a potential variable (Table \ref{tab:pos_variables}), and (b) the segments contain no verbs. This design is motivated by two critical observations. First, verbs are primary carriers of semantic content and are essential to a template's informativeness ~\cite{healy1970verb}. Second, unlike other parameters, verbs in logs exhibit limited morphological change and mostly function as constants. Merging segments with verbs would therefore sacrifice critical information.

\end{enumerate}

\begin{table}[tbp]
\centering
\scriptsize
\caption{Syntactic Roles with Their Relations to Potential Variables.}
\label{tab:pos_variables}
\begin{tabular}{|c|l|p{10cm}|}
\hline
\textbf{POS Tag} & \textbf{Examples} & \textbf{Rationale for Variable Classification} \\
\hline
X & \textit{@username}, \textit{URL} & 
Tokens labeled ``other/unknown'' (e.g., user-defined identifiers, URLs, emails, hashtags, domain-specific terms) that vary widely by context and cannot fit into constants. \\
\hline
NUM & \textit{42}, \textit{3.14} & 
Numerical values are inherently variable as they represent quantitative data that differs across instances. Numbers in templates typically correspond to user input, calculated results, or parametric values.  \\
\hline
PROPN & \textit{OpenAI}, \textit{Microsoft} & 
Proper nouns identify specific entities such as organizations, locations, or products. These are highly context-dependent and represent the most common type of variable content in templates, as they change based on the target domain or application. \\
\hline
\end{tabular}
\end{table}

Following the sequential merging assessment, templates are merged using the following methods, based on their segment difference. For segments with equal token numbers, we perform a fast merging method by a token-by-token comparison and replace differing tokens. For segments with unequal token counts, they are considered to have more complicated discrepancies, and an LLM will be used to merge the content semantically.
Each segment pair is processed independently. A complete template merge occurs if all segments qualify; otherwise, a partial merge is performed only on the qualifying segments, resulting in the creation of two new templates.



Upon completion of the merging process, whether producing one or two templates, the system removes the relevant templates from both the tree cache and bucket cache, subsequently inserting all newly generated templates into the cache structure.

\subsection{Correction Module Before Putting into Cache}

\begin{algorithm}[htbp]
\footnotesize
\renewcommand{\algorithmicrequire}{\textbf{Input:}}
\renewcommand{\algorithmicensure}{\textbf{Output:}}
\renewcommand{\algorithmiccomment}[1]{$\triangleright$ #1}
\caption{Sequential Template Validation}
\label{alg:correction_module_workflow}
\begin{algorithmic}[1]
\REQUIRE Log message $L_{ori}$, template $T_{ori}$
\ENSURE Validated template $T_{val}$

\COMMENT{Step 1: Format Correction}

\IF {$T_{ori}$ match $L_{ori}$}
    \STATE $T_{val} \leftarrow T_{ori}$
    \STATE goto Step 2
\ELSE
    \STATE $replaced\_parts \leftarrow$ DetectReplacedParts($L_{ori}$, $T_{ori}$)
    \STATE $T_{val} \leftarrow$ ReplaceWithPlaceholder($L_{ori}$, $replaced\_parts$)
\ENDIF

\COMMENT{Step 2: Over-specific Correction}

\STATE $special\_chars\_parts \leftarrow$ DetectRepeatedSpecialChars($T_{val}$)
\STATE $T_{val} \leftarrow$ CorrectOverSpecificByLLM($L_{ori}, T_{val}, special\_chars\_parts$)

\COMMENT{Step 3: Over-general Correction}

\STATE $has\_demonstration \leftarrow$ CheckDemonstration()
\IF{not $has\_demonstration$}
    \STATE $voc\_tokens \leftarrow$ ExtractVocabularyVariables($L_{ori}, T_{val}$)
    \STATE $T_{val} \leftarrow$ RestoreToConstant($T_{val}$, $voc\_tokens$)
\ELSE
    \STATE $pos\_verbs \leftarrow$ ExtractVerbsVariables($L_{ori}, T_{val}$)
    \STATE $T_{val} \leftarrow$ RestoreToConstant($T_{val}$, $pos\_verbs$)
\ENDIF

\end{algorithmic}
\end{algorithm}
The quality of LLM-generated templates is crucial for parsing accuracy, as these templates are cached for future reference. Therefore, to ensure a high-quality cache, we introduce a correction module as a ``quality guard'' to validate and correct templates before they are added.
Given a cache-missing log message $L_{ori}$ with its corresponding log template $T_{ori}$ generated by LLM, the correction module is designed to rectify potential template hallucination. There are three types of such errors: (1) format errors, where the template fails to match the input log; (2) over-specific errors, where variables are incorrectly classified as constants; and (3) over-general errors, where constants are incorrectly classified as variables.
Illustrative examples of these error types are presented in Table \ref{tab:template_errors}. 
The correction process includes three tailored error-handling components as illustrated in Algorithm \ref{alg:correction_module_workflow}, and we will detail these three sub-modules as follows.

\begin{table}[tbp]
\centering
\scriptsize
\caption{Examples of Template Error Types in Log Parsing}
\label{tab:template_errors}
\begin{tabular}{|p{2.2cm}|p{5cm}|p{5.5cm}|}
\hline
\textbf{Error Type} & \textbf{Correct Template} & \textbf{Erroneous Template} \\
\hline
(1) Format Error & 
\texttt{User <*> logged in from IP <*>} & 
\texttt{User <*> logged in{\color{red} ,} from IP <*>} \\
\hline
(2) Over-specific Error & 
\texttt{Process {\color{red} <*>} terminated with exit code <*>} & 
\texttt{Process {\color{red} apache2} terminated with exit code <*>} \\
\hline
(3) Over-general Error & 
\texttt{Database connection {\color{red} failed}} & 
\texttt{Database connection {\color{red} <*>}} \\
\hline
\end{tabular}
\end{table}

\subsubsection{Format Correction}
Smaller LLMs often struggle with parsing tasks due to their limited knowledge and weak understanding of context, which can lead to hallucinations ~\cite{zhang2025llm}. 
A common hallucination is bringing additional symbols like punctuation in parsing templates (Table \ref{tab:template_errors} (1)), leading to a mismatch using regular expressions. 
An intuitive solution is to re-query the LLM, which often yields the same incorrect output. However, this would be cost-inefficient, given that repeated LLM querying tends to reproduce identical errors.


Instead of using $T_{ori}$ directly, we reconstruct the template by leveraging the shared structure between $T_{ori}$ and $L_{ori}$. We begin by tokenizing $T_{ori}$ and $L_{ori}$ with a set of delimiters. Tokens present in $L_{ori}$ but absent from  $T_{ori}$ are identified as potential variables. Then, the final template is built by replacing these variable tokens with placeholder <*> in $L_{ori}$.
This ensures the resulting template is structurally identical to the original log.


\subsubsection{Over-specific Correction}
\begin{figure}[tbp]
    \vspace{2mm}
    \centering
    \includegraphics[width=1.0\linewidth]{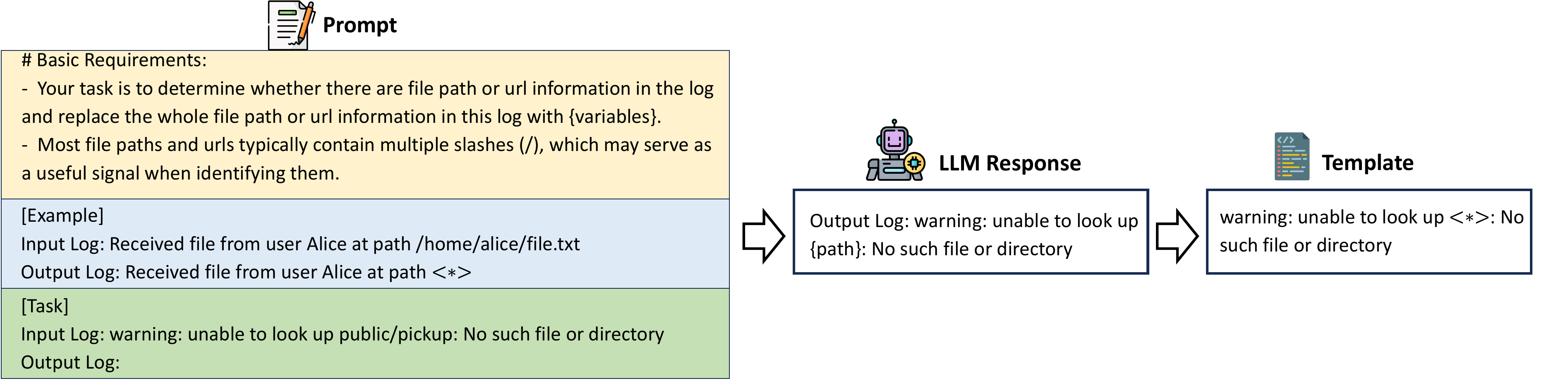}
    \caption{The design of over-specific correction prompts}
    \label{fig:over_specific_prompts}
\end{figure}

For long and complex logs, a smaller LLM may produce syntactically correct but over-specialized templates, failing to abstract all variables, which leads to the over-specific problem. To address this, we move beyond syntax and look into the semantic log variables (e.g., path, ID, user name), a concept shown to be informative for log parsing in prior work ~\cite{huo2023semparser, zhang2025logbase}.
First, we identify special characters in the template (e.g., slashes, symbols); if such a pattern is found, we proceed to the next step.
Then, we leverage LLMs to identify the variable semantics. In specific, we include the original log as input to the LLM, with specific instructions and examples for obtaining the suspected variable type (e.g., paths), as shown in Fig. \ref{fig:over_specific_prompts}. If the LLM abstracts the targeted segment in its output, we apply this correction to our template, resolving the over-specialization. This two-step process avoids the cost of full re-analysis every log message while correcting abstraction failures.

\subsubsection{Over-General Correction}
The third hallucination type is over-generalization, where the LLM incorrectly abstracts constants into variables. This is most prevalent when parsing novel log formats where no similar examples (demonstrations) are found for $L_{ori}$ in the template extraction module, forcing the model to rely solely on its own capabilities.
We have different strategies based on the availability of demonstrations. For no-demonstration cases, we check whether a token abstracted by the LLM exists in the spaCy English vocabulary ~\cite{vasiliev2020natural, schmitt2019replicable}. If so, we revert it to a constant. This effectively filters common words incorrectly flagged as variables. In the more common case where demonstrations are available, any abstracted tokens tagged as a verb are reverted to a constant, based on the observation that verbs in logs typically denote fixed events.

\subsection{Reference Log Exemplar Selector}

Given that acquiring a large number of labeled examples for supervised learning is labor-intensive in real-world scenarios, EFParser operates in an unsupervised, label-free setting.
Similar to established unsupervised parsers ~\cite{huang2025no, xiao2024free}, we construct a set of log exemplars (demonstrations) to guide the LLM. The selection process is designed to balance between: (1) high similarity between each exemplar and the target log, and (2) high diversity among the exemplars themselves.

Our selection process begins by identifying an initial pool of candidate logs. We use token-level Levenshtein distance to measure the similarity of each log in our corpus against the target log ~\cite{yujian2007normalized}, selecting any candidate that exceeds a predefined similarity threshold.
Next, these candidates are used to populate the final demonstration set, which is constrained to a fixed capacity. When the current demonstration set has not reached this capacity limit, candidate logs are added directly.  If the capacity is reached, 
the selector refines the set by incorporating new candidates that maximize overall diversity. This is achieved via a replacement strategy that substitutes an existing exemplar with a new candidate if the substitution results in a lower average intra-set similarity (i.e., token-level Levenshtein distance).




\section{Experiment Setup}

We evaluate EFParser by answering the following research questions (RQs):
\begin{itemize}[leftmargin=10pt, itemsep=0.5em]
    \item \textbf{RQ1:} How effective is EFParser?
    \item \textbf{RQ2:} What is the impact of each component on EFParser's performance?
    \item \textbf{RQ3:} How robust is EFParser across different configurations?
    \item \textbf{RQ4:} How efficient is EFParser?
\end{itemize}

\subsection{Datasets}

We conduct our experiments on Loghub-2.0 ~\cite{jiang2024large, he2020loghub}, a large-scale benchmark dataset for log parsing evaluation, aligning with existing studies ~\cite{jiang2024lilac,xiao2024free}. The dataset consists of 14 log collections from diverse system types, including distributed systems, supercomputers, operating systems, and server applications. Each collection in Loghub-2.0 is annotated with ground-truth templates and contains an average of 3.6 million log messages. The entire benchmark includes approximately 3,500 unique log templates, providing a comprehensive foundation for evaluating both parsing accuracy and efficiency.

\subsection{Baselines}
We select six leading log parsers for comparison, comprising three syntax-based and three semantic-based approaches. (1) Syntax-based: AEL ~\cite{jiang2008abstracting}, Brain ~\cite{yu2023brain} and Drain ~\cite{he2017drain}, which are chosen for their strong performance among syntax-based parsers. (2)  Semantic-based: we select LILAC ~\cite{jiang2024lilac}, LogBatcher ~\cite{xiao2024free}, and LUNAR ~\cite{huang2025no} because they are unsupervised like EFParser and thus better reflect practical deployment scenarios. 
LogBatcher and LUNAR sample logs that exhibit both commonality and variability with the target log as additional context to guide LLM-based unsupervised parsing. 
The original LILAC, by contrast, requires labeled examples for in-context learning. Following ~\cite{huang2025no}, to make LILAC comparable to the other unsupervised baselines, we replace its ICL module with 4-shot fixed demonstrations, producing the unsupervised variant denoted LILAC w/o ICL.
All baseline implementations were reproduced from their public repositories with their original parameters.

\subsection{Evaluation Metrics}
Following prior studies ~\cite{huang2025no,jiang2024lilac}, we adopt the five metrics for evaluating effectiveness and one for efficiency.

\begin{itemize}[leftmargin=10pt, itemsep=0.5em]
    \item \textbf{Grouping Accuracy (GA): } GA evaluates the capability to correctly group the log messages that belong to the same templates ~\cite{zhu2019tools}. It is computed by dividing the number of accurately grouped log messages by the total message count. A log message is accurately grouped if and only if its assigned template corresponds to the same set of log messages as indicated in the ground truth.
    \item \textbf{F1 score of Grouping Accuracy (FGA): } FGA focuses on the proportion of correctly grouped templates rather than correctly grouped log messages ~\cite{jiang2023large}, addressing the limitation of GA in scenarios where the majority of log messages (e.g., 95\%) belong to a small fraction of templates (e.g., 1\%). This template-level evaluation mitigates the high variance in GA that can occur due to the imbalanced distribution of messages across templates. FGA is computed as the harmonic mean of PGA (Precision of Group Accuracy) and RGA (Recall of Group Accuracy).
    \item \textbf{Parsing Accuracy (PA): } PA assesses the ability to correctly extract template parts and parameter parts from each log message ~\cite{dai2020logram}. It is computed as the ratio of correctly parsed log messages to the total number of log messages. A log message is deemed correctly parsed if and only if all static and variable tokens are accurately identified.
    \item \textbf{F1 score of Template Accuracy (FTA): } Similar to FGA, FTA also focuses on the template level by evaluating the accuracy of template identification ~\cite{khan2022guidelines}. FTA is computed as the harmonic mean of precision and recall of template accuracy. A template is deemed correct if and only if it satisfies two conditions simultaneously: (1) the corresponding log messages must belong to the same ground-truth template, and (2) the exact token sequence must match the ground-truth template precisely.
    \item \textbf{Processing Time: } Processing time refers to the total execution time, including preprocessing of raw datasets, LLM response generation, network latency, and post-processing of generated templates. Compared to the number of LLM invocations and token consumption, this end-to-end performance metric is a direct indicator of parser efficiency.
\end{itemize}

\subsection{Environment and Implementation}
All experiments are conducted on a MacBook Air equipped with an Apple M3 processor and 16GB of unified memory, running macOS Sequoia version 15.5. We employ Gemini-1.5-Flash-8B as the default LLM and invoke the LLM API through the OpenAI interface ~\cite{openai2023api}. To ensure deterministic results and eliminate randomness, the temperature parameter is set to 0. Additionally, we configure the default similarity threshold to 0.75 and limit the maximum number of demonstrations to 3.

\section{Evaluation Results}

\subsection{RQ1: How effective is EFParser?}
\begin{table}[tbp]
\centering
\caption{Comparison with the state-of-the-art log parsers, where Gemini-8b refers to Gemini 1.5 Flash-8B.\protect\footnotemark }
\label{effectiveness_result_table}
\begin{adjustbox}{max width = \textwidth}
\begin{tabular}{c|c|c|ccccccccccccccc}
\toprule
\textbf{Method} & \textbf{LLM} & \textbf{Metric} & Proxifier & Apache & OpenSSH & HDFS & OpenStack & HPC & Zookeeper & HApp & Hadoop & Spark & BGL & Linux & Mac & Thbd & \textbf{Average} \\
\toprule

\multirow{4}{*}{\textbf{Brain}} & {\cellcolor{gray!20}} & GA & 52.1 & \underline{99.7} & 66.3 & 96.0 & \textbf{100.0} & 80.0 & 99.3 & 97.9 & 56.3 & 97.2 & 94.0 & 79.0 & 83.4 & 79.2 & 84.3 \\
 & {\cellcolor{gray!20}} & PA & 70.3 & 28.7 & 48.1 & 92.9 & 14.1 & 66.3 & 82.2 & 17.5 & 14.3 & 39.3 & 40.2 & 1.0 & 32.5 & 26.1 & 41.6\\
 & {\cellcolor{gray!20}} & FGA & 73.7 & \underline{93.3} & 75.9 & 75.9 & \textbf{100.0} & 47.7 & 79.8 & 87.2 & 52.8 & 20.8 & 75.6 & 75.1 & 75.4 & 74.8 & 71.8\\
 & {\cellcolor{gray!20}} & FTA & 73.7 & 46.7 & 34.5 & 62.1 & 29.2 & 21.3 & 60.1 & 33.9 & 20.0 & 1.0 & 19.7 & 27.5 & 29.4 & 27.4 & 35.4\\
 \cline{1-1}\cline{3-18}

 \multirow{4}{*}{\textbf{Drain}} & {\cellcolor{gray!20}} & GA & 69.2 & \textbf{100.0} & \underline{70.7} & \underline{99.9} & \underline{75.2} & 79.3 & \underline{99.4} & 86.2 & 92.1 & 88.8 & 91.9 & 68.6 & 76.1 & \underline{83.1} & 84.3 \\
 & {\cellcolor{gray!20}} & PA & 68.8 & 72.7 & 58.6 & 62.1 & 2.9 & 72.1 & \underline{84.3} & 31.2 & 54.1 & 39.4 & 40.7 & 11.1 & 35.7 & 21.6 & 46.8\\
 & {\cellcolor{gray!20}} & FGA & 20.6 & \textbf{100.0} & \underline{87.2} & 93.5 & 0.7 & 30.9 & \textbf{\underline{90.4}} & 1.0 & 78.5 & \underline{86.1} & 62.4 & 77.8 & 22.9 & 23.7 & 55.4\\
 & {\cellcolor{gray!20}} & FTA & 17.6 & 51.7 & 48.7 & 60.9 & 0.2 & 15.2 & 61.4 & 0.4 & 38.4 & 41.2 & 19.3 & 25.9 & 6.9 & 7.1 & 28.2\\
 \cline{1-1}\cline{3-18}

 \multirow{4}{*}{\textbf{AEL}} & {\cellcolor{gray!20}} & GA & \underline{97.4} & \textbf{100.0} & 70.5 & \underline{99.9} & 74.3 & 74.8 & \textbf{99.6} & 72.5 & 82.3 & - & 91.5 & \textbf{91.6} & 79.7 & 78.6 & 85.6 \\
 & {\cellcolor{gray!20}} & PA & 67.7 & 72.7 & 36.4 & 62.1 & 2.9 & 74.1 & 84.2 & 31.1 & 53.5 & - & 40.6 & 8.2 & 24.5 & 16.3 & 44.2\\
 & {\cellcolor{gray!20}} & FGA & 66.7 & \textbf{100.0} & 68.9 & 76.4 & 68.2 & 20.1 & 78.8 & 0.8 & 11.7 & - & 58.7 & 80.6 & 79.3 & 11.6 & 55.5\\
 & {\cellcolor{gray!20}} & FTA & 41.7 & 51.7 & 33.3 & 56.2 & 16.5 & 13.6 & 46.5 & 0.3 & 5.8 & - & 16.5 & 21.7 & 20.5 & 3.5 & 25.2\\
 \cline{1-18}

\multirow{8}{*}{\textbf{LILAC w/o ICL}} & \multirow{4}{*}{GPT-3.5} & GA & 95.9 & \textbf{100.0} & 49.4 & \textbf{100.0} & \textbf{100.0} & \textbf{86.7} & 99.2 & \underline{99.3} & 91.5 & \textbf{98.1} & 87.9 & \underline{94.1} & 78.8 & 69.6 & 89.3 \\
 & & PA & \underline{95.9} & 96.5 & 62.6 & 91.5 & 48.5 & 93.6 & 43.7 & 56.2 & \textbf{85.1} &	99.1 & 91.5 & 75.4 & 54.1 & 53.4 & 74.8\\
 & & FGA & 66.7 & \textbf{100.0} & 69.7 & 65.8 & \textbf{100.0} & 79.1 & 89.2 & \underline{97.1} & 92.3 & \textbf{88.8} & \underline{85.0} & 88.1 & 76.3 & 33.1 & 80.8\\
 & & FTA & 66.7 & 62.1 & 54.5 & 52.1 & 83.3 & 67.6 & 71.1 & 74.3 & \underline{72.3} & \textbf{72.5} & 70.7 & 57.0 & 45.1 & 22.4 & 62.3\\
 \cline{2-18}
  & \cellcolor{gray!20}& \cellcolor{gray!20} GA & \cellcolor{gray!20}66.2 & \cellcolor{gray!20}92.6 & \cellcolor{gray!20}67.2 & \cellcolor{gray!20}\textbf{100.0} & \cellcolor{gray!20}50.5 & \cellcolor{gray!20}\textbf{87.0} & \cellcolor{gray!20}99.1 & \cellcolor{gray!20}\underline{99.3} & \cellcolor{gray!20}92.9 & \cellcolor{gray!20}\textbf{98.3} & \cellcolor{gray!20}79.7 & \cellcolor{gray!20}\underline{88.6} & \cellcolor{gray!20}75.5 & \cellcolor{gray!20}62.5 & \cellcolor{gray!20}82.8\\
 & \cellcolor{gray!20} & \cellcolor{gray!20}PA & \cellcolor{gray!20}70.6 & \cellcolor{gray!20}90.0 & \cellcolor{gray!20}11.2 & \cellcolor{gray!20}\underline{99.9} & \cellcolor{gray!20}44.5 & \cellcolor{gray!20}73.5 & \cellcolor{gray!20}49.2 & \cellcolor{gray!20}55.0 & \cellcolor{gray!20}72.8 & \cellcolor{gray!20}\textbf{98.1} & \cellcolor{gray!20}35.1 & \cellcolor{gray!20}\textbf{83.1} & \cellcolor{gray!20}44.2 & \cellcolor{gray!20}30.7 & \cellcolor{gray!20}61.3\\
 & \cellcolor{gray!20} & \cellcolor{gray!20}FGA & \cellcolor{gray!20}28.6 & \cellcolor{gray!20}70.4 & \cellcolor{gray!20}69.7 & \cellcolor{gray!20}65.8 & \cellcolor{gray!20}92.5 & \cellcolor{gray!20}83.5 & \cellcolor{gray!20}82.5 & \cellcolor{gray!20}95.2 & \cellcolor{gray!20}\underline{93.3} & \cellcolor{gray!20}83.7 & \cellcolor{gray!20}75.1 & \cellcolor{gray!20}74.4 & \cellcolor{gray!20}73.8 & \cellcolor{gray!20}28.2 & \cellcolor{gray!20}72.6\\
 & \cellcolor{gray!20} \multirow{-4}{*}{Gemini-8b} & \cellcolor{gray!20}FTA & \cellcolor{gray!20}38.1 & \cellcolor{gray!20}51.9 & \cellcolor{gray!20}21.2 & \cellcolor{gray!20}60.3 & \cellcolor{gray!20}77.4 & \cellcolor{gray!20}60.4 & \cellcolor{gray!20}66.2 & \cellcolor{gray!20}71.4 & \cellcolor{gray!20}66.5 & \cellcolor{gray!20}62.0 & \cellcolor{gray!20}52.9 & \cellcolor{gray!20}43.8 & \cellcolor{gray!20}38.1 & \cellcolor{gray!20}14.7 & \cellcolor{gray!20}51.8\\
 \cline{1-18}

\multirow{8}{*}{\textbf{LogBatcher}} & \multirow{4}{*}{GPT-3.5} & GA & 67.6 & \underline{99.7} & \underline{72.6} & \textbf{100.0} & 52.4 & 85.9 & 99.3 & \textbf{100.0} & 93.3 & \underline{97.3} & \underline{94.2} & \textbf{94.5} & 87.1 & 76.7 & 87.2\\
 & & PA & 67.6 & 99.1 & \textbf{83.6} & \underline{99.9} & 49.8 & 98.4 & \underline{96.8} & \textbf{97.7} & 70.5 & \textbf{99.6} & \underline{95.1} & \textbf{86.8} & 58.3 & \underline{57.6} & 82.9\\
 & & FGA & 72.7 & 92.1 & 84.0 & \textbf{100.0} & \underline{95.7} & \underline{86.5} & \textbf{96.1} & \textbf{98.1} & 86.9 & 79.0 & 84.5 & \underline{89.5} & 85.6 & 83.1 & 88.1\\
 & & FTA & 72.7 & 73.0 & 74.1 & \underline{95.7} & \underline{85.1} & \underline{83.9} & \textbf{87.2} & \textbf{85.9} & 64.2 & \underline{71.0} & \underline{75.5} & \underline{72.5} & 52.9 & \underline{59.0} & 75.2\\
 \cline{2-18}
 & \cellcolor{gray!20} & \cellcolor{gray!20}GA & \cellcolor{gray!20}70.3 & \cellcolor{gray!20}99.4 & \cellcolor{gray!20}\textbf{74.9} & \cellcolor{gray!20}\textbf{100.0} & \cellcolor{gray!20}\underline{96.0} & \cellcolor{gray!20}85.9 & \cellcolor{gray!20}98.7 & \cellcolor{gray!20}\textbf{99.8} & \cellcolor{gray!20}91.3 & \cellcolor{gray!20}94.7 & \cellcolor{gray!20}\underline{94.7} & \cellcolor{gray!20}82.2 & \cellcolor{gray!20}\underline{91.7} & \cellcolor{gray!20}85.3 & \cellcolor{gray!20}\underline{90.4}\\
 & \cellcolor{gray!20} & \cellcolor{gray!20}PA & \cellcolor{gray!20}\underline{75.1} & \cellcolor{gray!20}\underline{99.1} & \cellcolor{gray!20}\textbf{96.8} & \cellcolor{gray!20}76.2 & \cellcolor{gray!20}\underline{88.4} & \cellcolor{gray!20}\textbf{99.0} & \cellcolor{gray!20}82.3 & \cellcolor{gray!20}\textbf{97.2} & \cellcolor{gray!20}\textbf{85.4} & \cellcolor{gray!20}81.1 & \cellcolor{gray!20}\textbf{98.5} & \cellcolor{gray!20}\underline{80.7} & \cellcolor{gray!20}\textbf{64.8} & \cellcolor{gray!20}51.6 & \cellcolor{gray!20}\underline{84.0}\\
 & \cellcolor{gray!20} & \cellcolor{gray!20}FGA & \cellcolor{gray!20}\underline{76.2} & \cellcolor{gray!20}86.2 & \cellcolor{gray!20}\textbf{90.9} & \cellcolor{gray!20}\textbf{100.0} &\cellcolor{gray!20}\underline{96.8} & \cellcolor{gray!20}\underline{88.3} & \cellcolor{gray!20}\underline{89.9} & \cellcolor{gray!20}\textbf{98.1} & \cellcolor{gray!20}86.9 & \cellcolor{gray!20}78.8 & \cellcolor{gray!20}\underline{86.1} & \cellcolor{gray!20}\underline{83.7} & \cellcolor{gray!20}\textbf{89.4} & \cellcolor{gray!20}\underline{85.2} & \cellcolor{gray!20}\underline{88.3}\\
 & \cellcolor{gray!20} \multirow{-4}{*}{Gemini-8b} & \cellcolor{gray!20}FTA & \cellcolor{gray!20}76.2 & \cellcolor{gray!20}67.7 & \cellcolor{gray!20}\underline{64.9} & \cellcolor{gray!20}91.3 & \cellcolor{gray!20}\underline{84.2} & \cellcolor{gray!20}\textbf{90.9} & \cellcolor{gray!20}\textbf{80.5} & \cellcolor{gray!20}\underline{82.8} & \cellcolor{gray!20}69.6 & \cellcolor{gray!20}\underline{68.8} & \cellcolor{gray!20}\textbf{82.4} & \cellcolor{gray!20}\underline{69.7} & \cellcolor{gray!20}\textbf{58.0} & \cellcolor{gray!20}\textbf{63.0} & \cellcolor{gray!20}\underline{75.0}\\
 \cline{1-18}

 \multirow{8}{*}{\textbf{LUNAR}} & \multirow{4}{*}{GPT-3.5} & GA & \underline{98.9} & \textbf{100.0} & \textbf{78.0} & \textbf{100.0} & \textbf{100.0} & \underline{86.4} & 99.3 & \textbf{100.0} & \underline{94.1} & 95.0 & \textbf{95.5} & 83.0 & \underline{88.3} & \textbf{86.8} & \textbf{93.2}\\
 & & PA & \textbf{100.0} & \textbf{99.8} & \underline{72.2} & \textbf{100.0} & \underline{90.4} & \textbf{99.0} & 71.1 & \underline{96.6} & \underline{84.2} & 97.1 & \textbf{96.6} & 74.1 & \textbf{61.7} & \textbf{60.7} & \underline{86.0}\\
 & & FGA & \underline{87.0} & \textbf{100.0} & \textbf{92.3} & \underline{96.8} & \textbf{100.0} & 83.8 & 88.5 & \underline{97.1} & \underline{93.4} & \underline{87.9} & \textbf{87.3} & 87.4 & \underline{86.5} & \textbf{86.7} & \textbf{91.0}\\
 & & FTA & \textbf{95.7} & \textbf{86.2} & \textbf{92.3} & \textbf{96.8} & \textbf{89.6} & 82.6 & \underline{78.8} & \underline{85.0} & 69.3 & 64.4 & \textbf{79.2} & 71.5 & \underline{53.3} & \textbf{59.1} & \textbf{78.8}\\
 \cline{2-18}
 & \cellcolor{gray!20} & \cellcolor{gray!20}GA & \cellcolor{gray!20}50.9 & \cellcolor{gray!20}\underline{99.7} & \cellcolor{gray!20}70.4 & \cellcolor{gray!20}\textbf{100.0} & \cellcolor{gray!20}89.1 & \cellcolor{gray!20}\underline{86.4} & \cellcolor{gray!20}99.3 & \cellcolor{gray!20}\underline{99.3} & \cellcolor{gray!20}\textbf{95.2} & \cellcolor{gray!20}\underline{95.6} & \cellcolor{gray!20}\textbf{95.0} & \cellcolor{gray!20}82.7 & \cellcolor{gray!20}90.8 & \cellcolor{gray!20}\underline{86.6} & \cellcolor{gray!20}88.6\\
 & \cellcolor{gray!20} & \cellcolor{gray!20}PA & \cellcolor{gray!20}52.0 & \cellcolor{gray!20}40.1 & \cellcolor{gray!20}50.5 & \cellcolor{gray!20}\textbf{100.0} & \cellcolor{gray!20}80.4 & \cellcolor{gray!20}\textbf{99.0} & \cellcolor{gray!20}82.4 & \cellcolor{gray!20}\underline{96.5} & \cellcolor{gray!20}84.8 & \cellcolor{gray!20}88.3 & \cellcolor{gray!20}\underline{97.2} & \cellcolor{gray!20}77.7 & \cellcolor{gray!20}57.7 & \cellcolor{gray!20}\underline{56.3} & \cellcolor{gray!20}75.9\\
 & \cellcolor{gray!20} & \cellcolor{gray!20}FGA & \cellcolor{gray!20}72.7 & \cellcolor{gray!20}91.8 & \cellcolor{gray!20}78.4 & \cellcolor{gray!20}\textbf{100.0} & \cellcolor{gray!20}87.9 & \cellcolor{gray!20}83.3 & \cellcolor{gray!20}88.5 & \cellcolor{gray!20}\underline{95.9} & \cellcolor{gray!20}91.8 & \cellcolor{gray!20}\textbf{87.7} & \cellcolor{gray!20}\textbf{86.4} & \cellcolor{gray!20}82.8 & \cellcolor{gray!20}\underline{88.6} & \cellcolor{gray!20}\textbf{87.3} & \cellcolor{gray!20}87.4\\
 &  \cellcolor{gray!20}\multirow{-4}{*}{Gemini-8b} & \cellcolor{gray!20}FTA & \cellcolor{gray!20}\underline{81.8} & \cellcolor{gray!20}\underline{75.4} & \cellcolor{gray!20}40.5 & \cellcolor{gray!20}\textbf{100.0} & \cellcolor{gray!20}72.5 & \cellcolor{gray!20}83.3 & \cellcolor{gray!20}76.4 & \cellcolor{gray!20}\textbf{84.4} & \cellcolor{gray!20}\underline{70.1} & \cellcolor{gray!20}67.2 & \cellcolor{gray!20}\underline{79.6} & \cellcolor{gray!20}68.1 & \cellcolor{gray!20}\underline{57.1} & \cellcolor{gray!20}57.1 & \cellcolor{gray!20}72.4\\
 \cline{1-18}

 \multirow{8}{*}{\textbf{EFParser}} & \multirow{4}{*}{GPT-3.5} & GA & \textbf{100.0} & \underline{99.7} & 68.9 & \textbf{100.0} & \textbf{100.0} & 86.3 & 98.7 & 95.8 & \textbf{96.2} & 97.0 & 92.5 & 83.8 & \textbf{90.0} & 83.0 & \underline{92.3}\\
 & & PA & \textbf{100.0} & \underline{99.4} & 72.1 & \textbf{100.0} & \textbf{95.2} & \underline{98.9} & \textbf{98.2} & 90.7 & 84.0 & \underline{99.3} & 93.4 & \underline{83.4} & \underline{61.4} & 56.0 & \textbf{88.0}\\
 & & FGA & \textbf{100.0} & 91.8 & 82.1 & \underline{96.8} & \textbf{100.0} & \textbf{89.5} & 86.7 & 96.2 & \textbf{94.5} & 71.1 & 72.8 & \textbf{92.3} & \textbf{87.9} & \underline{83.2} & \underline{88.9}\\
 & & FTA & \underline{90.9} & \underline{75.4} & \underline{82.1} & \textbf{96.8} & \textbf{89.6} & \textbf{89.5} & 78.3 & 75.9 & \textbf{77.7} & 59.0 & 67.0 & \textbf{76.2} & \textbf{54.1} & 57.5 & \underline{76.4}\\
 \cline{2-18}
 & \cellcolor{gray!20} & \cellcolor{gray!20}GA & \cellcolor{gray!20}\textbf{100.0} & \cellcolor{gray!20}\underline{99.7} & \cellcolor{gray!20}61.8 & \cellcolor{gray!20}\textbf{100.0} & \cellcolor{gray!20}\textbf{100.0} & \cellcolor{gray!20}86.3 & \cellcolor{gray!20}\underline{99.4} & \cellcolor{gray!20}99.2 & \cellcolor{gray!20}\underline{95.0} & \cellcolor{gray!20}88.6 & \cellcolor{gray!20}94.3 & \cellcolor{gray!20}84.7 & \cellcolor{gray!20}\textbf{93.2} & \cellcolor{gray!20}\textbf{89.0} & \cellcolor{gray!20}\textbf{92.2}\\
 & \cellcolor{gray!20} & \cellcolor{gray!20}PA & \cellcolor{gray!20}\textbf{100.0} & \cellcolor{gray!20}\textbf{99.5} & \cellcolor{gray!20}\underline{64.7} & \cellcolor{gray!20}\textbf{100.0} & \cellcolor{gray!20}\textbf{98.2} & \cellcolor{gray!20}\underline{98.9} & \cellcolor{gray!20}\textbf{96.4} & \cellcolor{gray!20}95.6 & \cellcolor{gray!20}\underline{85.0} & \cellcolor{gray!20}\underline{95.1} & \cellcolor{gray!20}96.1 & \cellcolor{gray!20}80.0 & \cellcolor{gray!20}\underline{64.5} & \cellcolor{gray!20}\textbf{71.2} & \cellcolor{gray!20}\textbf{88.9}\\
 & \cellcolor{gray!20} & \cellcolor{gray!20}FGA & \cellcolor{gray!20}\textbf{100.0} & \cellcolor{gray!20}\underline{93.3} & \cellcolor{gray!20}80.5 & \cellcolor{gray!20}\underline{96.8} & \cellcolor{gray!20}\textbf{100.0} & \cellcolor{gray!20}\textbf{89.5} & \cellcolor{gray!20}89.2 & \cellcolor{gray!20}94.3 & \cellcolor{gray!20}\textbf{95.0} & \cellcolor{gray!20}79.6 & \cellcolor{gray!20}77.4 & \cellcolor{gray!20}\textbf{90.2} & \cellcolor{gray!20}88.5 & \cellcolor{gray!20}84.3 & \cellcolor{gray!20}\textbf{89.9}\\
 & \cellcolor{gray!20} \multirow{-4}{*}{Gemini-8b} & \cellcolor{gray!20}FTA & \cellcolor{gray!20}\textbf{90.9} & \cellcolor{gray!20}\textbf{80.0} & \cellcolor{gray!20}\textbf{77.9} & \cellcolor{gray!20}\underline{94.6} & \cellcolor{gray!20}\textbf{91.7} & \cellcolor{gray!20}\underline{89.5} & \cellcolor{gray!20}\underline{78.3} & \cellcolor{gray!20}75.8 & \cellcolor{gray!20}\textbf{79.2} & \cellcolor{gray!20}\textbf{69.4} & \cellcolor{gray!20}72.3 & \cellcolor{gray!20}\textbf{75.4} & \cellcolor{gray!20}54.2 & \cellcolor{gray!20}\underline{62.2} & \cellcolor{gray!20}\textbf{78.0}\\
 \cline{1-18}

\end{tabular}
\end{adjustbox}
\end{table}
\footnotetext{The symbol '-' indicates that the parsers failed to successfully parse the specific dataset due to various reasons, such as exceeding the predefined time limit of 12 hours. HApp and Thbd refer to HealthApp and Thunderbird, respectively.}

In this section, we conduct a comprehensive evaluation of EFParser's accuracy against both state-of-the-art syntactic-based and semantic-based log parsers. The results are presented in Table \ref{effectiveness_result_table}. The best performance for each metric is highlighted in bold, while the second-best performance is underlined. Grey background indicates using SLM as a backbone.

When evaluated on the smaller Gemini-1.5-8b model, EFParser demonstrates exceptional performance that significantly surpasses both traditional syntactic approaches and competing semantic-based methods. Compared to the best traditional syntactic parser (Brain), EFParser achieves remarkable improvements of 120.3\% in FTA (78.0\% vs. 35.4\%) and 25.2\% in FGA (89.9\% vs. 71.8\%), demonstrating the substantial advantages of semantic-based approaches. Among semantic parsers operating on the same small model, EFParser outperforms the competitive LUNAR method by 4.1\% in GA (92.2\% vs. 88.6\%) and 17.1\% in PA (88.9\% vs. 75.9\%). Most remarkably, EFParser's performance on the resource-efficient Gemini-1.5-8b model even surpasses some semantic parsers running on larger, more computationally expensive models, achieving 3.7\% higher FTA than LogBatcher on GPT-3.5 (78.0\% vs. 75.2\%) and 2.0\% higher FGA (89.9\% vs. 88.1\%), while closely matching LUNAR's GPT-3.5 performance with only a 1.0\% difference in FTA, thus demonstrating that EFParser can deliver near-optimal log parsing performance with significantly reduced computational requirements.

When evaluated on large-scale LLMs commonly employed by state-of-the-art semantic-based log parsers, EFParser demonstrates competitive performance that surpasses most existing methods across multiple evaluation metrics. Notably, EFParser achieves the highest PA score of 88.0\% among all parsers operating on GPT-3.5, representing a substantial improvement of 17.6\% over LILAC w/o ICL and 6.2\% over LogBatcher. While EFParser's performance on the remaining three metrics (GA, FGA, and FTA) averages only 1.8\% lower than LUNAR (currently the best-performing semantic-based log parser). This marginal difference can be attributed to the superior capabilities of large-scale LLMs. Specifically, as LLM capabilities increase, faulty template generation occurs less frequently, which paradoxically amplifies the adverse effects of overcorrection from EFParser's correction module and parsing cache mechanisms, resulting in slight performance degradation compared to simpler approaches that benefit more directly from enhanced model capabilities.

\begin{tcolorbox}[boxsep=1pt,left=2pt,right=2pt,top=3pt,bottom=2pt,width=\linewidth,colback=white!90!black,boxrule=0pt, colbacktitle=white!,toptitle=2pt,bottomtitle=1pt,opacitybacktitle=0]
\textbf{Answer to RQ1:} \textit{On small language models, EFParser consistently outperforms all baselines, including most of its LLM-based counterparts. This result underscores its effectiveness and practicality.}
\end{tcolorbox}

\subsection{RQ2: What is the impact of each component on EFParser's performance?}

\begin{table}[tbp]
\centering
\small
\caption{Ablation study of components of EFParser}
\label{tab:ablation_study}
\begin{tabular}{lcccc}
\toprule
& GA & PA & FGA & FTA \\

\midrule

EFParser & 92.2 & 88.9 & 89.9 & 78.0 \\

- w/o demonstrations & 84.5 ($\downarrow$8.4\%) & 79.5 ($\downarrow$10.6\%) & 84.4 ($\downarrow$6.1\%) & 72.9 ($\downarrow$6.5\%)\\

- w/o dual cache & 86.8 ($\downarrow$5.9\%) & 84.8 ($\downarrow$4.6\%) & 89.3 ($\downarrow$0.7\%) & 77.3 ($\downarrow$0.9\%) \\

- w/o format correction & - & - & - & - \\

- w/o over-specific correction & 92.2 & 82.3 ($\downarrow$7.4\%) & 89.7 ($\downarrow$0.2\%) & 74.7 ($\downarrow$4.2\%) \\

- w/o over-general correction & 91.0 ($\downarrow$1.3\%) & 87.1 ($\downarrow$2.0\%) & 87.6 ($\downarrow$2.6\%) & 74.3 ($\downarrow$4.7\%) \\
\bottomrule
\end{tabular}
\end{table}

\subsubsection{Settings}

In this research question, we evaluate the contribution of EFParser’s individual components: the dual cache, format correction, over-specific correction, and over-general correction. 
For evaluating the dual cache, we remove the bucket cache and adaptive updates, leaving only the tree cache and the same-length update strategy (matching prior semantic parsers). For evaluating the correction modules, each of the three correction modules is ablated individually to measure its impact on EFParser’s performance. The ablation study results are presented in Table \ref{tab:ablation_study}.


\subsubsection{Results.} The ablation study results are presented in Table \ref{tab:ablation_study}, with further analysis as follows.

\textit{Dual cache.} Removing the parsing cache leads to substantial performance degradation, with GA and PA decreasing by 5.9\% and 4.6\%, respectively. 
The reason for the performance drop is that, as the parsing cache merges semantically similar cached templates that should belong to the same template group into a unified representation. By merging similar templates, more log messages can be correctly grouped into their appropriate template categories, thereby improving GA. Additionally, the merging mechanism transforms erroneously identified constant tokens into appropriate wildcards, which enhances the accuracy of template structure and parameter extraction, leading to improved PA. The minimal change in FTA can be attributed to its stringent evaluation criteria, which require the token sequence of templates to exactly match the ground truth. Since parsing cache primarily focuses on template grouping rather than token-level refinement, it provides limited improvement in terms of precise token accuracy. Similarly, the negligible change in FGA stems from the fact that the majority of templates were already correctly grouped prior to cache implementation, resulting in a marginal effect due to the relatively small proportion of templates that benefit from the merging process.


\textit{Format correction.}  As shown in Table \ref{tab:ablation_study}, EFParser without the format correction module failed to complete processing on several of the 14 datasets within the designated one-hour time constraint. When operating with small LLMs, the system frequently generates templates that fail to match the input logs for certain log formats, resulting in excessive re-querying of identical log types. This inefficiency leads to computational overhead and increased token consumption. The practical impact of this limitation is demonstrated by specific datasets such as Thunderbird and OpenStack, which could not complete processing within the allocated time frame due to this iterative re-querying behavior, thereby confirming the critical importance of the format correction module for maintaining processing efficiency.

\textit{Over-specific correction.}
The result shows that removing the over-specific correction module significantly degrades performance: PA falls by 7.4\% and FTA by 4.2\%. This aligns with its role in detecting constants that were incorrectly classified as fixed tokens and converting them into variable placeholders. By restoring correct variable representations, the module enhances both template structure accuracy and parameter extraction for improving PA and FTA.

\textit{Over-general correction.}
Results in Table \ref{tab:ablation_study} show that the removal of the over-general correction module leads to a slight but consistent performance degradation across all evaluation metrics, with an average drop of approximately 2.5\%. This relatively modest impact, in contrast to the more substantial degradation in removing over-specific correction, indicates that the over-general errors rarely happen. Such faults predominantly occur in scenarios where the demonstration pool lacks appropriate examples for specific log patterns, a situation that arises with relatively low probability in practice.

\begin{tcolorbox}[boxsep=1pt,left=2pt,right=2pt,top=3pt,bottom=2pt,width=\linewidth,colback=white!90!black,boxrule=0pt, colbacktitle=white!,toptitle=2pt,bottomtitle=1pt,opacitybacktitle=0]
\textbf{Answer to RQ2:} \textit{The ablation study validates our design, showing all designed modules are effective. The parsing cache drives grouping accuracy at the message level, while the correction module refines fine-grained classifications at the token level.}
\end{tcolorbox}

\subsection{RQ3: How robust is EFParser across different configurations?}
\subsubsection{Robustness with Different LLMs.}
\begin{table}[tbp]
\centering
\small
\caption{EFParser's Performance with different backbones LLMs.}
\label{tab:different_llms}
\begin{tabular}{lccccc}
\toprule
Model & Parameters & GA & PA & FGA & FTA \\

\midrule
Gemini-1.5-flash & 8b & 92.2 & 88.9 & 89.9 & 78.0 \\

\midrule
LLaMA-3 & 8b & 92.5 ($\uparrow$0.3\%) & 86.8 ($\downarrow$2.4\%) & 89.5 ($\downarrow$0.4\%) & 74.9 ($\downarrow$4.0\%) \\

\midrule

\multirow{3}{*}{Qwen3} & 4b & 90.4 ($\downarrow$2.0\%) & 87.9 ($\downarrow$1.1\%) & 88.5 ($\downarrow$1.6\%) & 75.4 ($\downarrow$3.3\%)\\
 & 8b & 91.6 ($\downarrow$0.7\%) & 88.6 ($\downarrow$0.3\%) & 89.4 ($\downarrow$0.6\%) & 78.3 ($\uparrow$0.4\%) \\
 & 14b & 91.7 ($\downarrow$0.5\%) & 89.3 ($\uparrow$0.4\%) & 88.7 ($\downarrow$1.3\%) & 79.2 ($\uparrow$1.5\%) \\

\bottomrule
\end{tabular}
\end{table}
To evaluate EFParser's robustness across different LLM architectures, we tested five commonly used small-scale models: Gemini-1.5-flash-8b, LLaMA-3-8b-instruct, and three different-scale Qwen3 models. These model sizes span from 4b to 14b, representing typical small language models. As shown in Table \ref{tab:different_llms}, EFParser maintains consistently high performance across all models, with maximum deviation between average metrics limited to 2\%. This stability demonstrates EFParser's practical viability and reduces deployment constraints in real-world applications.

\subsubsection{Robustness with Different Similarity Thresholds.}
\begin{figure}[tbp]
    \vspace{2mm}
    \centering
    \includegraphics[width=1.0\linewidth]{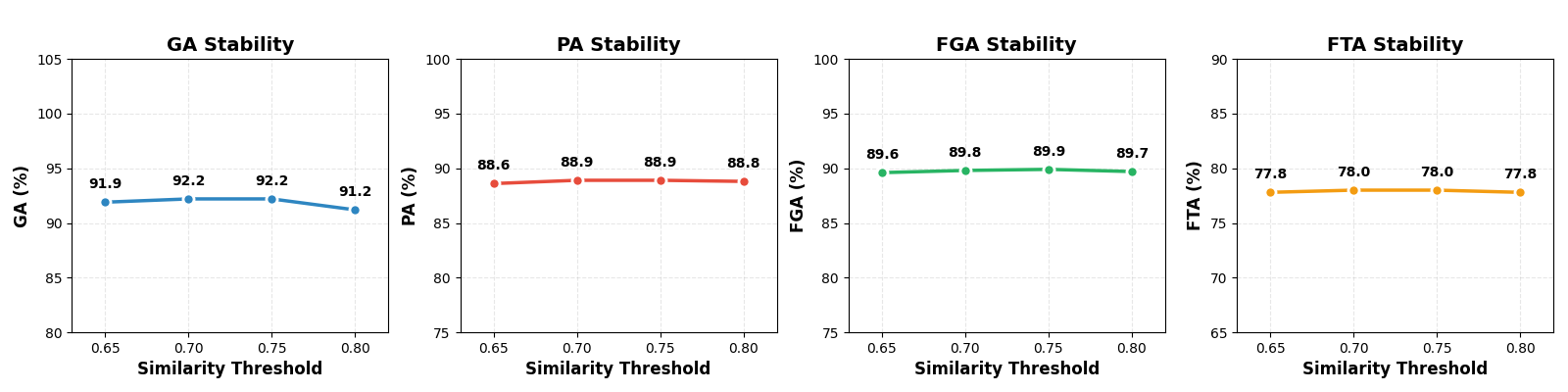}
    \caption{Performance with different similarity threshold}
    \label{fig:similarity_threshold}
\end{figure}

The similarity threshold serves as a critical parameter in EFParser, determining whether two templates exhibit sufficient similarity to warrant consolidation into a unified template. To evaluate the stability of EFParser's performance, it is essential to assess whether variations in the similarity threshold significantly impact overall system effectiveness. As demonstrated in Figure \ref{fig:similarity_threshold}, when the similarity threshold varies from 0.65 to 0.8, the fluctuation across all four performance metrics remains within 1.1\%. This finding validates that EFParser maintains consistently stable and high performance, demonstrating robustness across different similarity threshold configurations.

\begin{tcolorbox}[boxsep=1pt,left=2pt,right=2pt,top=3pt,bottom=2pt,width=\linewidth,colback=white!90!black,boxrule=0pt, colbacktitle=white!,toptitle=2pt,bottomtitle=1pt,opacitybacktitle=0]
\textbf{Answer to RQ3:} \textit{EFParser demonstrates strong robustness by achieving consistently high performance across different underlying LLM architecture, model size, and similarity threshold configurations.}
\end{tcolorbox}

\subsection{RQ4: How efficient is EFParser?}

\begin{figure}[tbp]
    \centering
    \includegraphics[width=0.8\linewidth]{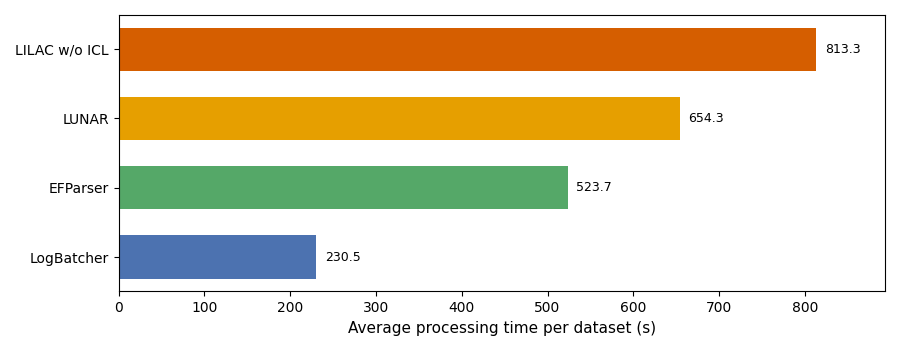}
    \caption{Processing Time Comparison of EFParser and Semantic-based Baselines on Loghub-2.0}
    \label{fig:efficiency_comparison}
\end{figure}

\subsubsection{Settings.} To ensure experimental fairness, we conduct efficiency comparisons between EFParser and baseline methods under controlled conditions. All experiments are performed on identical hardware configurations, with API calls executed through a unified interface to maintain consistency across different parsers. To mitigate the impact of occasional network latency variations, each experiment is replicated three times. Given the minimal variance observed across these repetitions, we report the averaged results as the final performance metrics.

\subsubsection{Results.} As illustrated in Figure \ref{fig:efficiency_comparison}, despite incorporating an additional cache updating strategy for new template insertion and a comprehensive correction module that validates generated templates through both manual-based and LLM-based mechanisms, EFParser demonstrates competitive processing efficiency. Notably, EFParser achieves 20\% and 35\% faster processing times compared to LUNAR and LILAC w/o ICL, respectively.

This performance advantage can be attributed to EFParser's superior template quality, which enables higher log-template matching success rates and reduces redundant processing. In contrast, LUNAR and LILAC w/o ICL suffer from consistently generating erroneous templates, leading to frequent template-log mismatches. Consequently, these methods require more re-querying operations when encountering previously processed but unsuccessfully matched logs, substantially degrading their overall processing efficiency.

\begin{tcolorbox}[boxsep=1pt,left=2pt,right=2pt,top=3pt,bottom=2pt,width=\linewidth,colback=white!90!black,boxrule=0pt, colbacktitle=white!,toptitle=2pt,bottomtitle=1pt,opacitybacktitle=0]
\textbf{Answer to RQ4:} \textit{Despite its additional cache updating and correction module, EFParser is more efficient than most semantic-based baselines.}
\end{tcolorbox}

\section{Threats to Validity}

\textbf{Internal threat.}
A potential threat to internal validity is the possibility of training data contamination, where the pre-trained LLM might have been exposed to our benchmark datasets. If so, EFParser's high performance could stem from memorization rather than true parsing capability.
To address this concern, we conduct an ablation study evaluating EFParser without demonstration examples from datasets. The results show significantly degraded performance compared to the full EFParser, suggesting that it is unlikely that the underlying LLMs memorize similar data. 


%

\noindent\textbf{External threat.}
Another threat is the inherent non-determinism of LLMs. Even with a temperature setting of 0, LLMs can produce stochastic outputs for identical inputs, which could introduce variance and bias into our performance comparisons.
To ensure the reliability and stability of our findings, we mitigated this threat by executing each experiment three times under identical conditions (temperature set to 0). The final reported performance metrics are the average results across three runs, providing a more stable and representative measure of each model's true capability.


\section{Related Work}

\subsection{LLM-Based Log Parser}

Log parsing is a critical prerequisite for automated log analysis tasks like anomaly detection and failure diagnosis. Recent methods leverage the powerful capabilities of LLMs for template extraction, offering superior generalization compared to syntactic-based methods that rely on statistical information and heuristic rules.
DivLog was the first LLM-based log parser ~\cite{xu2024divlog}, which samples representative logs and manually extracts their templates as references when querying the LLM. Building upon this, LILAC introduced a template cache mechanism that directly returns cached templates for known log patterns without LLM queries, significantly reducing time consumption and token usage ~\cite{jiang2024lilac}. LUNAR and LogBatcher further advanced the semantic-based log parsers by eliminating the requirement for manually labeled reference data ~\cite{huang2025no, xiao2024free}. For each log to be parsed, they dynamically sample logs from the dataset that exhibit both commonality and variability with the target log, enabling the LLM to learn constants and variables through comparison.

However, their performance is dependent on the capability of the backbone LLM, which degrades substantially when deployed with smaller, more resource-efficient models, limiting their practical applicability ~\cite{kaplan2020scaling}. Furthermore, prior work focuses on curating better prompts for LLMs, rather than improving the key caching strategy. To address this gap, we propose EFParser, a novel framework featuring a dual cache with an adaptive updating strategy and a self-correction module. This design ensures both high accuracy and efficiency across a wide range of LLM scales.

\subsection{Log Analysis}

Log analysis utilizes log parser outputs, templates, and parameter sequences to extract insights from system logs. Templates represent structural patterns, while parameters capture dynamic values, collectively forming the foundation for various downstream analysis tasks. Among these applications, anomaly detection serves as one of the most representative tasks. Current anomaly detection methods can be mainly divided into two types: forecasting-based (unsupervised) and classification-based (supervised). 

For forecasting-based methods, DeepLog pioneered the use of LSTM models for log anomaly detection by learning normal execution patterns and predicting the next log event ~\cite{du2017deeplog}. Anomalies are detected when the actual log event deviates from the prediction. LogAnomaly extends this forecasting approach by introducing template2vec, which represents log templates as semantic vectors to better match new log events with existing templates during the forecasting process ~\cite{meng2019loganomaly}.

In contrast to forecasting-based methods, classification-based approaches directly classify log sequences as normal or abnormal rather than predicting future events. LogRobust leverages a pre-trained Word2vec model and combines it with TF-IDF weights to capture token importance, feeding these representations into an Attention-based Bi-LSTM for binary classification ~\cite{zhang2019robust}. CNN-based methods transform log template sequence into trainable matrices, using convolutional operations to detect anomalous patterns ~\cite{lu2018detecting}. NeuralLog ~\cite{le2021log} applies a Transformer-based architecture for its strong semantic understanding ability to directly detect anomalies without parsing.

\section{Conclusion}
In this paper, we propose EFParser, an unsupervised LLM-based log parser designed to address the performance degradation when existing semantic-based parsers are transferred to smaller LLMs due to their framework design deficiencies. EFParser incorporates a comprehensive framework with two key components: a dual cache module with adaptive updating mechanism that merges similar templates to maintain consistency and a correction module that validates and refines LLM-generated templates. Experimental results on large-scale datasets demonstrate that EFParser significantly outperforms existing baseline methods on smaller LLMs, with performance even exceeding some methods using large-scale LLMs. The combination of superior accuracy and computational efficiency makes EFParser highly suitable for real-world deployment where data privacy concerns require local model deployment. This work demonstrates that systematic architectural design can effectively mitigate the limitations of smaller LLMs in log parsing tasks.

\section*{Data Availability}
All the code for EFParser is available at https://github.com/LogAnalysisTech/EFParser-Log-Parser for replication and future research \cite{wang_2026_19280693}.

\section*{Acknowledgment}
This research was supported by the Google Academic Research Award and Singapore Ministry of Education (MOE) Academic Research Fund (AcRF) Tier 1 grant (Project ID: 23-SIS-SMU-088). Any opinions, findings and conclusions or recommendations expressed in this material are those of the author(s) and do not reflect the views of the Ministry of Education, Singapore.

\bibliographystyle{ACM-Reference-Format}
\bibliography{references}

\end{document}